\documentstyle[preprint,aps,psfig,12pt]{revtex}
\begin{document}
\preprint{\begin{tabular}{l}
\hbox to \hsize{2000 June \hfill BROWN-HET-1231}
\end{tabular}}
\bigskip
\def\gsim{\ \rlap{\raise 3pt \hbox{$?$}}{\lower 3pt \hbox{$\sim$}}\ }
\def\lsim{\ \rlap{\raise 3pt \hbox{$<$}}{\lower 3pt \hbox{$\sim$}}\ }

\title
{Gravitational Origin of Quark Masses and Mixings in an Extra-Dimensional Brane-World}
\author{David Dooling$^{a}$ and Kyungsik Kang$^{a}$}

\address{a. Department of Physics\\
Brown University, Providence RI 02912, USA}
\maketitle
\tightenlines
\begin{abstract}
Using the resolution of the gauge hierarchy problem recently proposed by Randall and Sundrum, we find a natural explanation for the observed fermion masses and mixings of the three Standard Model (SM) generations.
Localizing massless SM matter generations on neighboring 3-branes in an extra dimensional world leads to effective four dimensional masses and mixings from the coupling of the fermions with the background metric.
We find that the positions of the 3-branes required to solve the gauge hierarchy problem simultaneously reproduces phenomenologically acceptable fermion masses and mixings.
\end{abstract}

\bigskip
\centerline {\rm\bf I. Introduction}

The Standard Model (SM) provides an elegant mechanism by which the massive intermediate vector bosons W$^{\pm}$ and Z acquire mass while the photon and gluons remain massless.
Postulating the Higgs field to transform as a singlet under SU(3)$_{c}$ and a doublet under SU(2), the W$^{\pm}$ and Z masses at tree level are given in terms of $g_{1}$, $g_{2}$ and only one dimensional parameter $v \sim$ 246 GeV.
The SM does not provide such an appealing and economical explanation for the observed fermion masses.
After spontaneous symmetry breaking, the fermion mass term in the Lagrangian reads:

\begin{equation}
\mathcal{L}_{\mathnormal{mass}} \mathnormal = \frac{v}{\sqrt{2}} \left( \overline{u}_{Li} h_{ij}^{(u)} u_{Rj} + \overline{d}_{Li} h_{ij}^{(d)} d_{Rj} \right) + h.c.
\end{equation}
where the $h_{ij}$ are arbitrary $3 \times 3$ complex Yukawa coupling matrices.
Some predictive mechanism for fermion mass generation is needed so as to place the understanding of fermion mass on a par with that of gauge boson mass; i.e., to find an overriding principle that predicts the values of $h_{ij}$ to be what they are.

Because the $h_{ij}$ are arbitrary, the flavor eigenstates do not necessarily coincide with the mass eigenstates.
When expressed in terms of mass eigenstates, the charged-current weak interactions involving left-handed quarks will involve different combinations of flavor eigenstates and one must introduce four additional physically meaningful parameters, the three angles and single phase of the quark mixing matrix.
This proliferation of free parameters severely limits the predictive power of the SM.
A first modest step is to have the four mixing parameters under control, as in the phenomenological realization of calculability; i.e., the idea of making ans\"{a}tze for the mass matrices such that all flavor-mixing parameters depend solely on the quark masses themselves \cite{1}.
The second step is to then have the $h_{ij}$ themselves determined by some as yet not understood mechanism involving even fewer parameters.
Recently, the introduction of large extra dimensions has been used to address some of the outstanding problems in our understanding of fundamental physics \cite{2}.
The idea of localizing different fermions in different places in an extra dimension as an ingredient in the explanation of the fermion mass hierarchy is due to Arkani-Hamed and Schmaltz, while mass hierarchies for fermions in Randall-Sundrum models have also been discussed recently by Gherghetta and Pomarol \cite{3}.
In this paper, we investigate the possibility that fermions do not acquire mass via Yukawa couplings with a Higgs scalar at all, but rather through coupling with gravity in a higher-dimensional brane world.
The elements of both the effective four dimensional up-quark and down-quark mass matrices are completely determined by this coupling as well as the positions of the respective branes in the extra infinite fifth dimension.
Requiring that the Randall-Sundrum resolution of the outstanding gauge hierarchy problem is operative in our scenario fixes the positions of the branes \cite{4}.
Hence the effective four dimensional mass matrices are completely determined modulo the widths of the respective guassian profiles of the localized generations, as will be discussed in the following sections.
Imposing strong calculability; i.e., all the widths to be the same, and also requiring trustworthiness of the solution (small brane widths with respect to the fundamental mass scale $M$ of the R-S scenario), we remarkably reproduce acceptable quark masses and mixings.

The outline of this paper is as follows.
In section I, we briefly summarize the Randall-Sundrum brane world and its infinite-dimensional extension of Lykken and Randall \cite{4}.
In section II, we motivate placing different SM generations on different branes, or at slighly differing coordinates within a single brane as pioneered by Dvali and Shifman \cite{5}.
In section III, we summarize and apply some of the formalism developed by Grossman and Neubert for neutrinos to the SM quarks and compute the now determined up-quark and down-quark type quark mass matrices and their eigenvalues \cite{6}.
Finally, we state our intriguing results and discuss the quark flavor mixing parameters.

\section{R-S Brane World and its $\infty$ Extra-Dimension Extension}

The particular brane-world geometry we will use in this paper is that proposed by Lykken and Randall \cite{4}.
This scenario resolves the gauge hierarchy problem by localizing the graviton away from the 3+1 - dimensional world where the SM resides, and because of the small amplitude for the graviton to coincide with our brane, the large value of the Planck scale arises.
The solution of Einstein's equations for the geometry of a single brane with cosmological energy densities tuned to guarantee Poincare invariance leads to the non-factorizable metric

\begin{equation}
ds^{2} = e^{-2k|y|} \eta_{\mu \nu} dx^{\mu} dx^{\nu} - dy^{2}
\end{equation}
where $\mu, \nu$ parametrize the four-dimensional coordinate of our world and $y$ is the coordinate of the fifth dimension.

If one introduces a compact, orbifold geometry, then one can have a positive energy brane at one point and a negative energy brane at the second orbifold point and a slice of AdS$_{5}$ space in between.
Lykken and Randall have proposed a scenario where negative tension branes are not required, and in which it is consistent to live with an infinite fifth dimension while still resolving the gauge hierarchy problem in a consistent manner.
One of the advantages of this extension of the orginal work of Randall and Sundrum is that now one can exploit the known mechanisms for confining gauge and matter fields on positive energy D-branes.
As in \cite{4}, our setup is a Planck brane on which the graviton zero mode is confined, exponentially falling off in the direction $y$.
We also have multiple branes located at a distance $\sim$ $y_{0}$ from this brane, where e$^{-ky_{0}} = \mathnormal \frac{TeV}{M_{PL}}$, where $k$ is related to the cosmological constant on the brane and determines the exponential falloff of the graviton.
The new branes can be regarded as a probe of the geometry determined by the Planck brane, either by assuming that the Planck brane has much larger tension, or consists of a large set of branes.
We will work in the above infinite five dimensional scenario with a single Planck brane located at the origin in the extra dimension and multiple postive tension ``probe'' branes at neighboring points in the extra dimension.
Our work is largely inspired by the recent papers by Dvali and Shifman \cite{5}, and Grossman and Neubert \cite{6}.

\section{Families as Neighbors}

In \cite{5}, Dvali and Shifman have used a five dimensional scenario where the fermion mass hierarchy is generated by a nonzero Higgs profile in the bulk which generates masses for the SM fermions localized on other branes.
Because in this framework the families are literally neighbors in the extra space, the well known pattern of ``nearest neighbor mixing'' results.
The pioneering idea we adopt from \cite{3,5} is that the hierarchical fermion masses arise from a geometrical origin rather than from some spontaneously broken flavor symmetry.
Just as in \cite{5}, ``We assume that three SM families are identical, the difference in their masses is simply because they happen to live in different places in the extra space.
More precisely, we assume that the original higher dimensional theory admits, as its solution, a brane with localized fermions with quantum numbers of one SM generation.
Multiple brane states will then generate $\nu$ identical copies of fermions, $\nu$ generations of the Standard Model''.

The crucial difference between the current investigation of families as neighbors in extra dimensions and that originally proposed in \cite{5}, is that we not couple the fermion matter generations to Higgs scalars at all.
Electroweak symmetry breaking (ESWB) occurs on each SM brane separately to give the appropiate gauge bosons mass, but is not responsible at all for generating fermion mass.
But an attractive feature of \cite{5} that we retain in the current work is the inevitable correlation between masses and mixings.
As will be described in the next section, we will place heavier generations closer to the Planck brane and successively lighter generations farther out in the extra dimension.
Assuming a gaussian profile around each brane and that all of the SM fermions are generated from a single progenitor family in the original five-dimensional theory, we can therefore identify branes at coordinates $y_{1}, y_{2}, y_{3}, y_{4}, y_{5}$ and $y_{6}$ with the flavor states $t, b, c, s, u$ and $d$.
We can consider the separation of SU(2) doublets as a result of EWSB on each of three orginal SM branes so that there are really three branes, but with SU(2) partners displaced within each brane.
As in \cite{5}, the finite distance between the branes means that the flavor eigenstates are not completely orthogonal.
There is a nonzero overlap between the wave functions and a nontrivial correlation between the fermion mixing and their masses, to be described in detail in the following section.

\section{Gravitational Origin of Quark Mass Matrices}

In \cite{6}, the original compact, orbifold Randall-Sundrum geometry with two branes located at the orbifold fixed points was exploited along with bulk fermion fields to derive a strong hierarchy of neutrino masses and large mixing angles.
As stressed in \cite{6}, the resolution of the gauge hiarchy problem in this scenario eliminates the standard see-saw mechanism as a viable model for neutrino mass and mixing, since the highest energy scale governing physics on the visible brane is now the weak scale.
But a higher dimensional treatment of fermions in the orginal Randall-Sundrum scenario can still provide a solution to the puzzling neutrino anamolies in the following way.
The extension of the Dirac algebra to five dimensions leads to a different propagation of left- and right- handed modes.
Localization of a right-handed zero mode on the hidden brane provides a new mechanism for obtaining small neutrino masses, which arise from coupling the Higgs and left-handed lepton fields of the SM, localized on the visible brane, to a right-handed fermion in the bulk.

In this work, we will use the infinite dimensional Randall-Sundrum brane world with no Higgs-fermion coupling to construct effective four dimensional quark mass matrices.
The action for a Dirac fermion propagating in a five-dimensional space with the metric Eq. (2). can be written in the form \cite{6,7}

\begin{equation}
S = \int d^{4}x \int dy \sqrt{G} \left\{ E^{A}_{a} \left[ \frac{i}{2} \overline{\Psi} \gamma^{a} \left( \partial_{A} - \stackrel{\leftarrow}{\partial_{A}} \right) \Psi \right] \right\} 
\end{equation}
where $G = \det{G_{AB}} = e^{-8k|y|}$ is the determinant of the metric.
We use capital indices $A, B,$... for objects defined in curved space, and lower-case indices for objects defined in the tangent frame.
The matrices $\gamma^{a} = \left( \gamma^{\mu}, i \gamma_{5} \right)$ provide a four-dimensional representation of the Dirac matrices in five-dimensional flat space.
The quantity $E_{a}^{A} = \mbox{diag} \left( e^{k|y|}, e^{k|y|}, e^{k|y|}, e^{k|y|}, 1 \right)$ is the inverse vielbein.

After integrating by parts and defining left and right handed spinors $\Psi_{L,R} \equiv \frac{1}{2} \left( 1 \mp \gamma_{5} \right) \Psi$, the action can be written as:

\begin{eqnarray}
S &=& \int\!\mbox{d}^4x\!\int\!\mbox{d},y\,\bigg\{
e^{-3k|y|} \left( \bar\Psi_L\,i\rlap/\partial\,\Psi_L
+ \bar\Psi_R\,i\rlap/\partial\,\Psi_R \right) \nonumber\\
&&\mbox{}-\frac{1}{2} \left[ \bar\Psi_L \left(
e^{-4k|y|} \partial_y + \partial_y\,e^{-4k|y|}
\right) \Psi_R - \bar\Psi_R \left(
e^{-4k|y|} \partial_y + \partial_y\,e^{-4k|y|}
 \Psi_L \right) \right] \bigg\} \,,
\end{eqnarray}
where we assume the fields fall off to zero at $\pm \infty$ in the $y$ direction.

With the above action, we see that an effective four dimensional fermion mass term can be generated by integrating over the $y$ dimension.
As noted above, the extension of the Dirac algebra to five dimensions results in different solutions for left- and right-handed fermions.
Our model is a five-dimensional effective theory of some fundamental ten dimensional string theory or eleven dimensional M theory, but we would like to see some manifestation of this fundamental theory in our five dimensional model.
Presumabably, in the fundamental theory, the graviton field and the SM matter and gauge fields are all members of some higher dimensional multiplet. 
We encode this relation by assuming the following solutions for $\Psi_{L}$ and $\Psi_{R}$ \cite{6}:

\begin{equation}
\Psi_{L} \sim \cos (ky) \left( \sum_{i=1}^{6} e^{-\frac{1}{2}k_{f}^{2}(y-y_{i})^{2}} \right)
\end{equation}
\begin{equation}
\Psi_{R} \sim \sin (ky) \left( \sum_{i=1}^{6} e^{-\frac{1}{2}k_{f}^{2}(y-y_{i})^{2}}  \right)
\end{equation}
i.e., we assume that the SM matter fields are described by guassian profiles localized around the branes with coordinates $y_{i} (i=1,2,3,4,5,6)$.
The trigonometric functions in front have the same $k$ in their arguments as appears in the warp factor of the metric and serves as the vestigial signature of the more fundamental theory, and $k_{f}$ is the inverse width of the gaussian profiles and is chosen to be identical for all of the SM branes.
As mentioned above, $y_{1}$ and $y_{2}$ will represent one SM brane, with the top quark, being the most massive, localized around $y_{1}$, and its SU(2) partner the bottom quark localized around $y_{2}$.
The complete dictionary between the coordinate in the fifth dimension and flavor is then given by $(y_{1}, y_{2}, y_{3}, y_{4}, y_{5}, y_{6} ) \rightarrow (t,, b, c, s, u, d)$.
Before constructing the quark mass matrices we must normalize the overall wave function as well as choose the $y_{i}$ in such a manner that the gauge hierarchy problem is addressed.
We define \cite{8}


\begin{equation}
n = \int_{-\infty}^{\infty} e^{-3k|y|} \left( \sum_{i=1}^{6} e^{-\frac{1}{2}k_{f}^{2}(y-y_{i})^{2}} \right)^{2}
\end{equation}

As introduced in \cite{4}, an important aspect of this geometry is the correspondence between location in the fifth dimension and the overall mass scale.
From a four dimensional perspective, the warp factor $e^{-k|y_{i}|}$ is a conformal factor.
Masses on the $i^{th}$ brane are rescaled by $e^{-k|y_{i}|}$, so that a natural scale for mass parameters might be $M_{pl} = 10^{19}$ GeV on the Planck brane at the origin, but $M_{pl}e^{-k|y_{i}|}$ for physics located at the $i^{th}$ SM brane.
The hierarchy between the TeV and the Planck scales can be successfully explained by an exponential of order 30, a dimensionless number.

The above metric can be trusted as a solution to the above geometry provided $k_{f} \ll k \sim M_{pl} \sim 10^{19}$ GeV.
Our strategy for fixing the $y_{i}$ so that the gauge hierarchy problem is resolved and that a numerical determination of the quark mass matrices is possible is as follows.
First we define
\begin{equation}
\Psi_{iL} = \frac{1}{\sqrt{n}} \cos (ky) e^{-\frac{1}{2}k_{f}^{2}(y-y_{i})^{2}}
\end{equation}
\begin{equation}
\Psi_{jR} = \frac{1}{\sqrt{n}} \sin (ky) e^{-\frac{1}{2}k_{f}^{2}(y-y_{j})^{2}}.
\end{equation}
Then using Eq.(4), we can construct the matrix
\begin{equation}
M_{ij} = \frac{1}{2n} \int_{-\infty}^{\infty} dy e^{-4k|y|} \left( \Psi_{iL} \left( \partial_{y} \Psi_{jR} \right) - \left( \partial_{y} \Psi_{iL} \right) \Psi_{jR} \right)
\end{equation}
The effective four dimensional up-type and down-type quark mass matrices will then be:
\begin{equation}
M_{u} = \left( \begin{array}{ccc} 
M_{55} & M_{53} & M_{51} \\
M_{35} & M_{33} & M_{31} \\
M_{15} & M_{13} & M_{11} 
\end{array} \right)
, M_{d} = \left( \begin{array}{ccc} 
M_{66} & M_{64} & M_{62} \\
M_{46} & M_{44} & M_{42} \\
M_{26} & M_{24} & M_{22} 
\end{array} \right)
\end{equation}
These quark mass matrices are not symmetric, and so to get the proper mass eigenvalues we form $\mathcal{M_{\mathnormal{u}}} = \mathnormal M_{u} M_{u}^{\dagger}$ and $\mathcal{M_{\mathnormal{d}}} = \mathnormal M_{d} M_{d}^{\dagger}$.

Because only the dimensionless product $ky_{i}$ is determined if we demand a taming of the gauge hierarchy problem, we are free to choose $k$ to be whatever we want provided $k \sim M_{pl}$.
The simplest choice is $k = 1$, with the units to be determined in a consistent fashion as described below.
We then choose $y_{1} = 37, y_{2} = 37.6, y_{3} = 38.2, y_{4} = 38.8, y_{5} = 39.4, y_{6} = 40$, where the choice of units will be explained below.
With this set of coordinates for the SM branes, the corresponding mass scales range from $M_{pl} e^{-40} \sim 42$ GeV to $M_{pl} e^{-37} \sim 850$ GeV.
Because in our scenario, different flavors are localized at different positions in the $y$-direction and are thus associated with different characteristic energy scales, and also because $\mathcal{M_{\mathnormal{u}}}$ and $\mathcal{M_{\mathnormal{d}}}$ result from coupling different flavors at different positions, we will compare our resulting mass eigenvalues with the running quark masses as in \cite{9}.

\bigskip
\bigskip
\vglue -1cm
\hglue -1cm
\psfig{figure=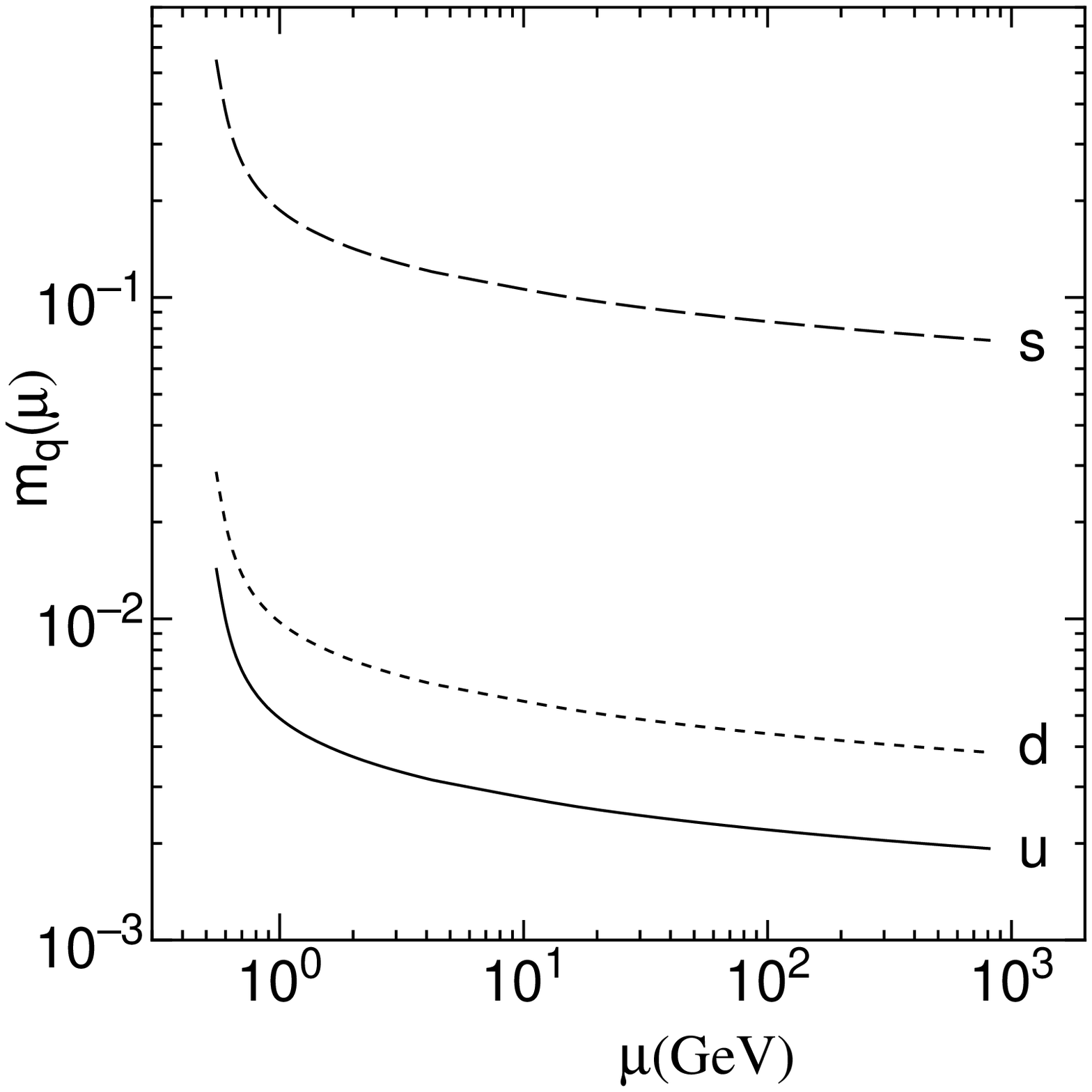,height=7cm,angle=0}\hglue 1cm
\psfig{figure=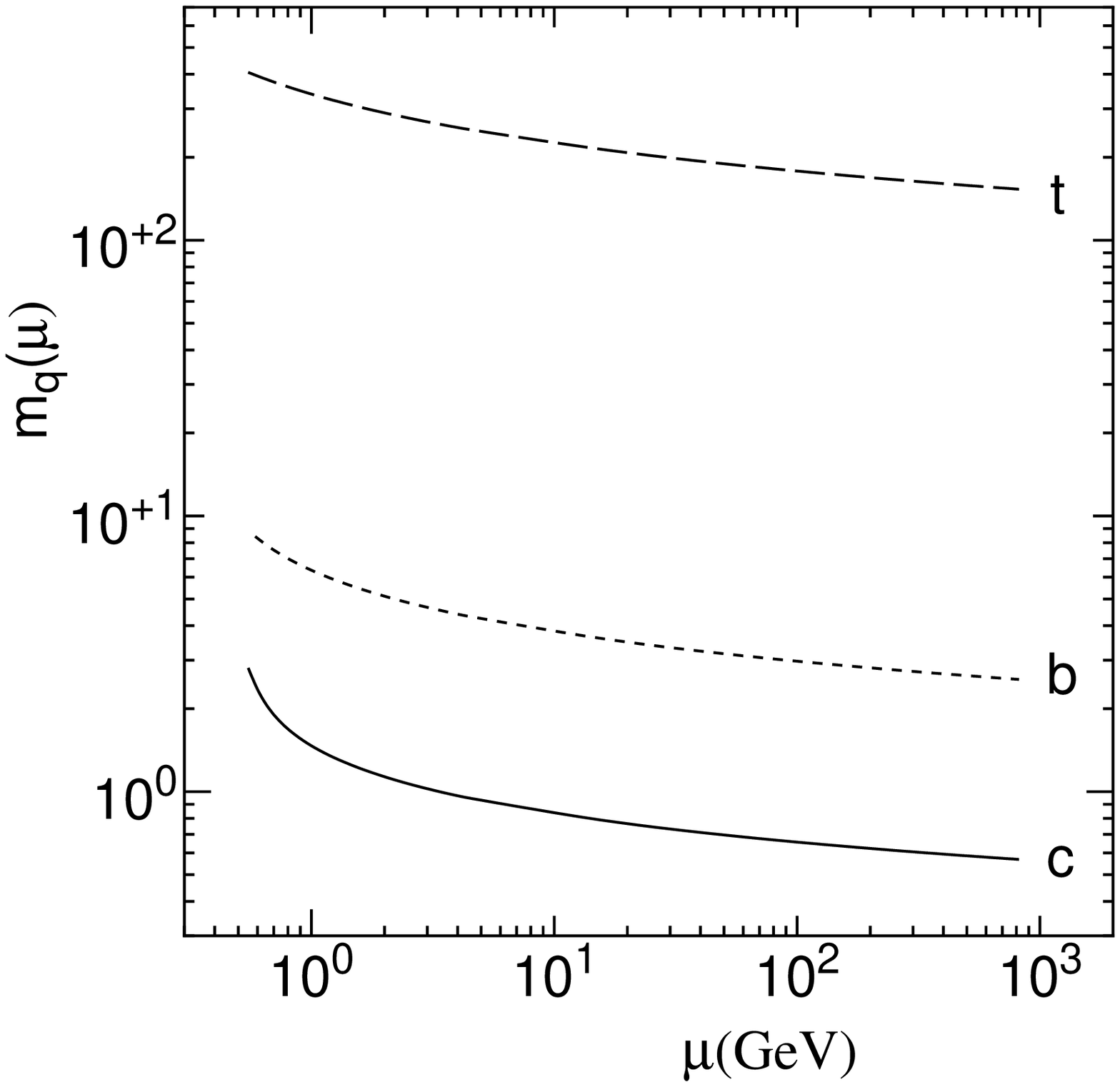,height=7cm,angle=0}
\normalsize

As mentioned, the resolution of the gauge hierarchy problem only determines the products $k|y_{i}|$.
Having made the choice $k=1$, we can determine the actual units by ``calibrating'' in the following manner.
First we choose a specific value for $k_{f}$, keeping in mind the requirement that $k_{f} < k \sim M_{pl}$.
Choosing for example $k_{f} = 0.293$, we arrive at the following numerical values for $\mathcal{M_{\mathnormal{u}}}$ and $\mathcal{M_{\mathnormal{d}}}$:

\begin{equation}
\mathcal{M_{\mathnormal{u}}} = \left( \begin{array}{lll}
.1522148233 \times 10^{-40} & .2006278629 \times 10^{-39} & .1846456179 \times 10^{-38} \\
.2006278629 \times 10^{-39} & .26477721456 \times 10^{-38} & .2431703393 \times 10^{-37} \\
.1846456179 \times 10^{-38} & .2431703393 \times 10^{-37} & .2241250654 \times 10^{-36} \end{array} \right)
\end{equation}
\begin{equation}
\mathcal{M_{\mathnormal{d}}} = \left( \begin{array}{lll}
.9643041321 \times 10^{-43} & .1347251462 \times 10^{-41} & .2297379411 \times 10^{-41} \\
.1347251462 \times 10^{-41} & .1894384118 \times 10^{-40} & .3073779455 \times 10^{-40} \\
.2297379411 \times 10^{-41} & .3073779455 \times 10^{-40} & .7024382435 \times 10^{-40} \end{array} \right)
\end{equation}
The numerical eigenvalues of $\mathcal{M_{\mathnormal{u}}}$ are computed to be $.3627082319 \times 10^{-45}, .9280804731 \times 10^{-41}$ and $.2267787272 \times 10^{-36}$.
We now determine the units by setting the heaviest eigenvalue equal to a typical value of the running top quark mass within the relevant energy range, say 210 GeV.
This means that the above eigenvalues with dimension mass-squared are measured in units of $x = \frac{210}{\sqrt{.2267787272 \times 10^{-36}}} GeV$, so that the warp factor $k \sim .4409792333 \times 10^{21} \sim M_{pl}$, consistent with the requirements of the Randall-Sundrum scenario.
Notice that a remarkable thing has happened above; we have used the experimental value of the top quark mass to determine the actual units of $k$, one distinct dimensionful ingredient of the dimensionless warp factor.
Even though we localized the flavor top quark on a brane with warp factor $e^{-k|y_{1}|}$ so as to address the gauge hierarchy problem, coordinates in the y-direction only entered in a complicated manner to determine the eigenvalues of $\mathcal{M_{\mathnormal{u}}}$ and $\mathcal{M_{\mathnormal{d}}}$.
We use the top quark mass directly to determine the units of $k$ and find that $k \sim M_{pl}$, implying strongly that the Randall-Sundrum resolution of the gauge hierarchy problem may simultaneously shed an equally bright light onto the mystery of the quark mass hierarchy.

Continuing to compute the quark spectrum in the above example, we find $m_{c} \sim \frac{x}{\sqrt{.9280804731 \times 10^{-41}}} \sim 1.34$ GeV, $m_{u} \sim \frac{x}{\sqrt{.3627082319 \times 10^{-45}}} \sim .008$ GeV, $m_{b} \sim \frac{x}{\sqrt{.8471110434 \times 10^{-40}}} \sim 4.05$ GeV, $m_{s} \sim \frac{x}{\sqrt{.4572984136 \times 10^{-41}}} \sim .94$ GeV and $m_{d} \sim \frac{x}{\sqrt{.7471251201 \times 10^{-47}}} \sim .001$ GeV.
Inspection of Figures (1) and (2) shows that all of these values are in good to excellent agreement with the known values of the running quark masses in the relevant energy range as determined by the warp factors associated with the SM branes.
With the exception of the strange quark, all of the determined masses lie well within the acceptable range.
The higher-dimensional determination of $m_{s}$ is off by approximately a factor of two, which disagreement is not necessarily an indication of a failing of our model, as determination of $m_{s}$ is a notoriously difficult enterprise, given that the strange quark is really neither a heavy nor a light quark.

We are now ready to discuss the quark flavor mixing that results from our model.
Forming $V_{CKM}$ from $U_{u}^{\dagger}U_{d}$, where $U_{u}^{\dagger}\mathcal{M_{\mathnormal{u}}}\mathnormal U_{u} = \mbox{diag} (m_{u}^{2}, m_{c}^{2}, m_{t}^{2} )$ and $U_{d}^{\dagger} \mathcal{M_{\mathnormal{d}}} \mathnormal  U_{d} = \mbox{diag} (m_{d}^{2}, m_{s}^{2}, m_{b}^{2} )$ from the above example gives:

\begin{equation}
V_{CKM} = \left( \begin{array}{ccc} 
.9995170807 & -.02787344213 & .01373595897 \\
.03082193824 & .9454946927 & -.3241755572 \\
-.003951387670 & .3244423762 & .9458972099 \end{array} \right)
\end{equation}
Clearly, the above $V_{CKM}$ does not agree with those found in the Particle Data Group report.
The experimental values of the $V_{CKM}$ elements are determined from weak decays of quarks, or, in some cases, from deep inelastic neutrino scattering \cite{10}.
Most of the experimental determinations of the elements correspond to an energy scale of approximately 1 GeV.
But the $V_{CKM}$ elements run with renormalization scale and the above mixing matrix does not correspond to a renormalization point of 1 GeV, but rather to the range of energy scales $M_{pl}e^{-k|y_{6}|}$ to $M_{pl}e^{-k|y_{1}|}$ \cite{9,11}.
From the familiar perspective of our four dimesional SM, we certainly have the freedom to perform a biunitary transformation on the above quark mass matrices that leaves both the quark mass spectrum and the flavor-mixing parameters unchanged, where the new mass matrices in both the up-type and down-type quark sectors are of the nearest neighbor interactions (NNI) form \cite{12}.
In \cite{13}, it is shown that starting from the NNI texture form, a very simple ans\"{a}tze leads to strong calculability; i.e., all of the flavor mixing parameters being functions solely of the quark masses.
In \cite{13}, masses evaluated at a modified minimal subtraction $(\overline{MS})$ renormalization point of 1 GeV were used and able to reproduce the observed values of the flavor mixing parameters.
If calculability is indeed a feature of nature, then the fact that the above mass matrices lead to the correct masses at the appropriate energy ensures that the correct flavor mixing information is also encoded in our model.
We do not simply move the branes further out in the $y$ direction so as to correspond to an energy scale of 1 GeV primarily because this new geometry would no longer address the gauge hierarchy with the same degree of success.

In conclusion, we have found a consistent, straitforward way within the extra dimensional scenario to derive effective four dimensional mass matrices that are phenomenologically acceptable.
We find that the positioning of the copies of SM 3-branes in the extra dimension that leads to a quark mass spectrum in good to excellent agreement with experiment is precisely that which also resolves the gauge hierarchy problem. 
Thus we have found very suggestive evidence that the solution to the gauge hierarchy problem also may provide the long sought-after overriding principle by which the elements of the quark mass matrices may be determined.

\section{Acknowledgements}

We wish to thank Mihail Mihailescu and Radu Tatar for helpful discussions.
D.D. wishes to thank the U. S. Department of Education for financial support via the Graduate Assistance in Areas of National Need (GAANN) program.
Support for this work was provided in part by U.S. Dept. of Energy grant DE-FG02-91ER40688 and an NSF Dissertation Enhancement Award (INT-0083352).


\begin{thebibliography}{99}

\bibitem{1} S. Weinberg, in: L. Motz(Ed.), a Festschrift for I.I. Rabi, NY Academyu of Sciences, New York, 1977; A.C. Rothman, K. Kang, Phys. Rev. Lett. 43 (1979) 1548; Phys. Rev. D 24 (1981) 167; H. Fritzsch, Phys. Lett. B 73 (1978) 317; Nucl. Phys. B 155 (1979) 189; A. De Rujula, H. Georgi, S.L. Glashow, Ann. Phys. (NY) 109 (1977) 258; F. Wilczek, A. Zee, Phys. Lett. B 79 (1977) 418; H. Georgi, D.V. Nanapoulos, Nucl. Phys. B 155 (1979) 52.
\bibitem{2} I. Antoniadis, Phys. Lett. B 264 (1990) 377; I. Antoniadis, C. Munoz, M. Quiros, Nucl. Phys. B 397 (1993) 515; I. Antoniadis, N. Arkani-Hamed, S. Dimopoulos, G. Dvali, Phys. Lett. B 436 (1998) 257; K.R. Dienes, E. Dudas, T. Ghergetta, Phys. Lett. B 436 (1998) 55; Nucl. Phys. B 537 (1999) 47; N. Arkani-Hamed, S. Dimopoulos, G. Dvali, Phys. Lett. B 429 (1998) 263; H. Hatanaka, T. Inami, C. S. Lim, Mod. Phys. Lett. A 13 (1998) 2601; K. Yoshioka, Mod. Phys. Lett. A 15 (2000) 29; D. Kaplan, T. Tait, JHEP 0006:020 (2000).
\bibitem{3} N. Arkani-Hamed, M. Schmaltz, Phys. Rev. D 61 (2000) 033005; T. Gherghetta, A. Pomarol , e-print hep-ph/0003129.
\bibitem{4} L. Randall, R. Sundrum, Phys. Rev. Lett. 83 (1999) 3370; J. Lykken, L. Randall, JHEP 0006:014,2000.
\bibitem{5} G. Dvali, M. Shifman, Phys. Lett. B 475 (2000) 295.
\bibitem{6} Y. Grossman, M. Neubert, Phys. Lett. B 474 (2000) 361.
\bibitem{7} T. Eguchi, P.B. Gilkey, A.J. Hanson, Phys. Rep. 66 (1980) 213; R.A. Bertlmann, \underline{Anomalies in Quantum Field Theory}, Oxford University Press, 1996.
\bibitem{8} V. Balasubramanian, P. Kraus, A. Lawrence, Phys. Rev. D 59 (1999) 046003.
\bibitem{9} H. Fusaoka, Y. Koide, Phys. Rev. D 57 (1998) 3986.
\bibitem{10} Particle Data Group, Caso et al., The European Physical Journal C 3 (1998) 1.
\bibitem{11} A. Barroso, L. Br\"{u}cher, R. Santos, hep-ph/00041362.
\bibitem{12} G.C. Branco, L. Lavoura, F. Mota, Phys. Rev. D 39 (1989) 3443.
\bibitem{13} D. Dooling, K. Kang, Phys. Lett. B 455 (1999) 264.
\end{thebibliography}
\end{document}